# The Geometric-Optical Mechanism of Wavefront Reconstruction

A.M. Smolovich

*Kotelnikov Institute of Radioengineering and Electronics (IRE) of Russian Academy of Sciences, Moscow, Russia*
*e-mail: asmolovich@petersmol.ru*

**Abstract**–The mechanism of wavefront reconstruction by a geometric-optical reflection of reconstructing light from surfaces with constant phase differences between the object and reference waves used to record the interference fringe structure in the medium bulk is analyzed. This mechanism is compared with the holographic one. Experiments in volume medium and in planar optical waveguide were carried out. Several types of achromatic optical elements are proposed. The optoacoustic focusing element, which can create a great local stress in a special shaped area inside the matter, is discussed. The possibility of the ultrashort pulse temporal reconstruction is demonstrated.

## INTRODUCTION

The mechanism of wavefront reconstruction by a hologram is based on the diffraction of reconstructing light by the recorded quasiperiodic interference fringe structure. Just the local fringe period of this structure contains the information about the object wavefront. This mechanism works both for thin and volume holograms and for holograms recorded both in transmitting and reflecting geometries. According to another mechanism suggested in [1] the object wavefront is reconstructed due to a geometric-optical (GO) reflection of reconstructing radiation from a single interference fringe maximum surface, which operates like a mirror with complicated curvature. Its local curvature contains the information about the object wavefront. In contrast to holography during the GO reconstruction there is no reconstructing light diffraction by periodic interference structure. This makes the reconstruction process an achromatic one, i.e. it permits to reconstruct an undistorted wavefront with light whose wavelength is different to that used in the recording process [2, 3]. The GO mechanism of wavefront reconstruction was suggested for holography. However, it differs significantly from the holographic mechanism and is of independent interest.

In this paper the GO mechanism is compared with the holographic one. We explain why the GO mechanism does not work in common holograms including Denisyuk holograms. Ultrashort laser pulse registration in a thick recording medium is used for the GO wavefront reconstruction. In the other option, a 2D analog of the effect is obtained in a planar optical waveguide. Several types of achromatic optical and optoacoustical elements (OE) and systems are proposed. Additionally, ultrashort pulse temporal reconstruction is discussed.

## PRINCIPLES OF GO RECONSTRUCTION

Let the object $A_O(\mathbf{r}) \exp[ikL_O(\mathbf{r})]$ and reference $A_R \exp[ikL_R(\mathbf{r})]$ waves satisfy the scalar equations of GO [4]. Here $k=2\pi/\lambda$ is the wave number, and $\lambda$ is the wavelength of light, $L_O(\mathbf{r})$ and $L_O(\mathbf{r})$ are the eikonals of the waves, $A_O(\mathbf{r})$ and $A_R$ are the amplitudes, $\mathbf{r}$ is the position vector. The intensity of the interference field of these waves is given by the expression:

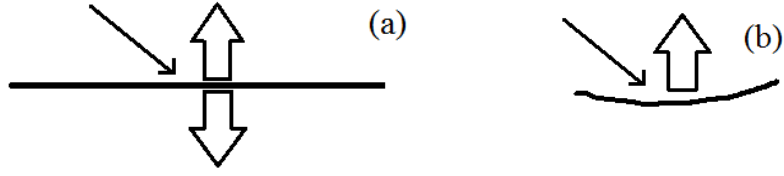

Figure 1. Wavefront reconstruction: (a) by hologram, (b) by GO mechanism.

$$A_O^2(\mathbf{r}) + A_R^2 + 2A_O(\mathbf{r})A_R \cos\{k[L_R(\mathbf{r}) - L_O(\mathbf{r})]\}. \quad (1)$$

First, let us consider the case when a thin hologram is recorded (figure 1a). Let the hologram be reconstructed by the wave $A_R \exp[ik`L_R(\mathbf{r})]$ where the wave number $k`=2\pi/\lambda`$ may differ from the wave number during the hologram recording. The resultant field produced by the interaction (as a result of transmission or reflection) of the reconstructing wave with the hologram can be expressed from (1) as a sum, where the term associated with the reconstructed wavefront is proportional to [5]:

$$A_O(\mathbf{r})A_R^2 \exp\{i[(k` - k)L_R(\mathbf{r}) + kL_O(\mathbf{r})]\}. \quad (2)$$

The eikonal of wave (2) is:

$$L_{diffr}(\mathbf{r}) = (\lambda`/\lambda) L_O(\mathbf{r}) + (1 - \lambda`/\lambda) L_R(\mathbf{r}). \quad (3)$$

It follows that the eikonal of the object wave will be reconstructed only if $\lambda` = \lambda$. For arbitrary $\lambda`$, (3) describes dispersion and for the case of plane waves leads to the grating formula [6], while for spherical waves it gives the Meir formula [7].

Following [1], we now consider surfaces on which the argument of cosine in (1) is constant (figure 1b). For brevity we will call them isophase surfaces, although, more precisely, they are surfaces on which the difference of the phases of the interfering waves is constant, or surfaces on which the intensity of the total field is constant. These surfaces are given by the equation:

$$L_R(\mathbf{r}) - L_O(\mathbf{r}) = p, \quad (4)$$

where $p$ is a constant for each isophase surface. If the reconstructing wave $A_R \exp[ik`L_R(\mathbf{r})]$ is GO reflected from the surface (4), then the phase $k` L_{refl}(\mathbf{r})$ of the reflected wave on this surface coincides with the phase of incident wave [4]:

$$k` L_{refl}(\mathbf{r}) = k` L_R(\mathbf{r}). \quad (5)$$

From (4) and (5) we obtain:

$$L_{refl}(\mathbf{r}) = L_O(\mathbf{r}) + p, \quad (6)$$

i.e., the eikonal of the object wave is reconstructed up to an additive constant for arbitrary $k`$, in contrast to the case of a thin hologram. Expression (6) also implies that the wavefront surface $L_O(r) = const$ is achromatically reconstructed. So, the waves reconstructed by these two mechanisms are different for $\lambda` \neq \lambda$.

The following simple examples show the difference between holographic reconstruction and GO reconstruction. In the simplest case of a plane object and reference waves, holographic reconstruction corresponds to reconstructing wave diffraction by a grating, and GO reconstruction corresponds to the reconstructing wave reflection by a plane mirror. If the object wave is spherical, corresponding structures are a Fresnel zone plate and a parabolic mirror. In general, GO reconstruction is similar to the reflection of a mirror with a complicated curvature.

An ordinary volume hologram contains domains of interference fringe maxima surfaces. These constitute a periodic structure containing a large number of such surfaces. As a rule, a volume hologram is recorded in a flat slab with a thickness between a few and tens of microns and with a lateral size between a few and tens of centimeters. So, the length of a periodic structure along the hologram surface usually exceeds by several orders the length of each part of an interference fringe maximum surface inside the slab (figure 2a). As a result, the diffraction mechanism of reconstruction dominates upon the GO mechanism [8, 9]. This is confirmed by the presence of dispersion during the reconstruction process of ordinary volume holograms that can be experimentally observed within the spectral selectivity band.

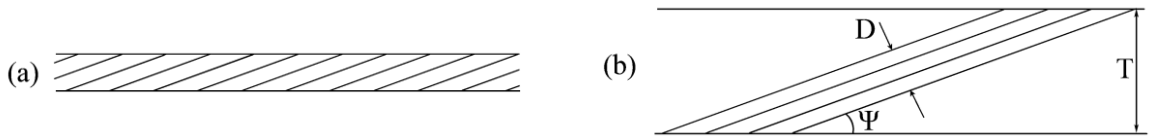

Figure 2. Interference fringe maxima surfaces in: (a) common volume hologram, (b) structure recorded by ultrashort pulses in medium with increased thickness.

## EXPERIMENTAL

To obtain GO reconstruction, the number of interference maxima recorded in the bulk must be decreased and the thickness of recording medium should be increased. For this, recording of ultrashort laser pulse interference can be used [10, 11]. The following criterion indicates the condition for GO mechanism domination for the hologram recorded on a flat slab:

$$T\cos\psi \gg D, \qquad (7)$$

where $T$ is the flat slab thickness, $\psi$ is an angle between the grating vector and the normal to the slab and $D$ is a width of the interference fringe structure in the direction normal to isophase surfaces (figure 2b). If the hologram was recorded by two short pulses with the duration equal to $\tau$ having plane wavefronts and propagating in the straight opposite directions $D \sim \tau c$, where $c$ is the speed of light inside the recording medium. The criterion (7) was obtained in [8, 9] by using the classical Gabor-Stroke approach [12].

We used femtosecond-pulse sapphire titanate laser with the radiation wavelength varying in the range of 780-830 nm for recording. The pulse duration was about 40 fs. The pulse repetition frequency was equal to 80 MHz with average power of 200mW. The laser beam was divided with the aid of a half-transmitting interference mirror into two beams, which were directed toward one another. The recording plate was positioned so that its photosensitive layer was located in the region of overlap of the oppositely propagating pulses. The method proposed in [13] was used to precisely equalize the lengths of the optical paths. The normal of the plate made an angle on the order of 20° with the optical axis. Specially prepared photographic plates with a photosensitive layer of thickness from 110 to 280 μ were used. The exposure was set experimentally and was on the order of a tenth of a Joule per square centimeter. Several fields with different exposure were irradiated on each photographic plate. As a control, some fields were exposed with continuous radiation in the same geometry as with the pulsed radiation. The reconstruction was done in reflection geometry. An argon-laser pumped dye laser (rhodamine 6G solution in ethylene glycol) was used for reconstruction. A continuous lasing regime with wavelength tuning from 580 to 630 nm with the aid of a dispersing element was used. The reconstructed beam was observed on a screen placed about 4 m from the photographic plate. The reconstructed beam was observed during variation of the light wavelength from 585 to 607 nm. With this wavelength variation the spot produced on the screen by the reconstructed beam did not move. Contrary to this, the spot from the beam reconstructed by the hologram recorded with the continuous radiation under the same conditions moved by 6 cm. So, we can draw a conclusion that the GO mechanism of reconstruction takes place for structure recorded by femtosecond pulses.

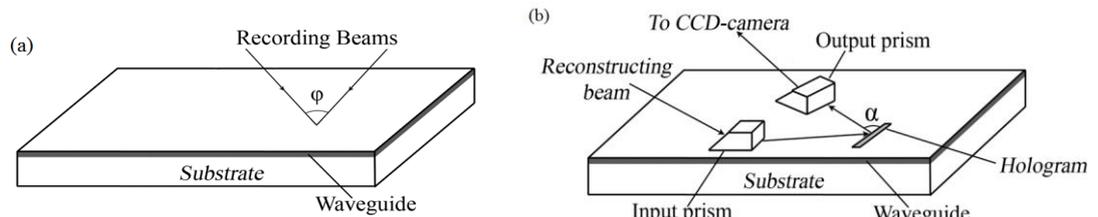

Figure 3. GO mechanism in planar waveguide: (a) recording, (b) reconstruction.

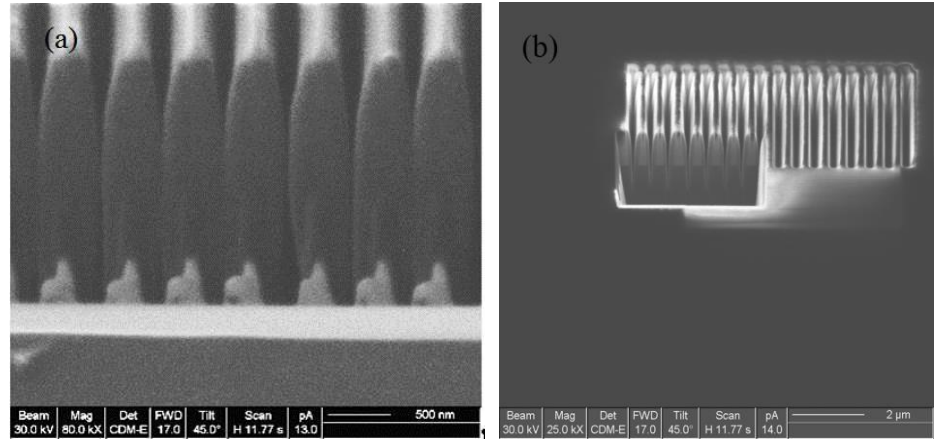

Figure 4. Grating fabricated by FIB: (a) on polymer, (b) on silicon.

However, the large thickness of the recording medium required for the effect to be observed is a rather serious obstacle for its application. It is difficult to treat such thick media due to obvious technical difficulties. Also, in such thick slabs the absorption during recording [14] leads to a great decrease of the diffraction efficiency [15, 16]. To overcome this drawback we propose to use the waveguide analog of the GO reconstruction effect. Suppose that the hologram is recorded by ultrashort laser pulses that enter the planar waveguide through its upper surface (figure 3a). The reconstruction occurs in the waveguide regime (figure 3b). In this case the reconstruction process is two-dimensional. The top view of the area of waveguide hologram lit by reconstructing beam is similar to the cross section in the 3D case (figure 2b). Therefore, we can use criterion (7) of GO reconstruction replacing the slab thickness $T$ for the reconstructing beam width in waveguide plane. Here the interference fringe structure width $D \sim \tau c/\sin(\phi/2)$, where $\phi$ is the angle between the beams recording the hologram. It can be seen that now the GO reconstruction exhibits itself irrelevantly on the waveguide thickness. This is a major advantage of the waveguide GO reconstruction over its three-dimensional counterpart.

In the experiment we used a 20 μm film of dichromated gelatin ($n_2 = 1.54$) on a polished quartz substrate ($n_1 = 1.456$) as a planar waveguide model. The second harmonic (400 nm) of the Ti: Sapphire laser pulses were used for recording (figure 3a). The pulses duration was about 30 fs. The pulses were split and directed into the waveguide sample through its upper surface. The angle between the recording beams was approximately 60°. The exposed dichromated gelatin films were soaked in cold water and dried in a sequence of isopropyl alcohol solutions baths of increasing concentrations. The hologram reconstruction was done in waveguide regime (figure 3b). We used the right angle prisms made of Yttrium Aluminum Garnet ($Y_3Al_5O_{12}$) with a 60° angle at the contact side for the beam input and output. A dye laser, with Rhodamine 6G as the dye, pumped with the second harmonic (532nm) of a Nd:YAG laser was used for reconstruction. The wavelength was varied from 580 to 630 nm. The output signal was observed within the band of 580-615 nm corresponding to the hologram spectral selectivity. The direction of the output beam was measured with an accuracy of about $6 \times 10^{-4}$ rad. For the holographic reconstruction mechanism, it should be dependent on the reconstructing wavelength. The whole deviation of the calculated direction of the holographically reconstructed output beam within the hologram spectral selectivity band would be more than 0.03 rad. The direction of the output beam observed in the experiment did not change. This is proof that the GO mechanism of wavefront reconstruction takes place here.

Instead of direct holographic recording the waveguide achromatic hologram containing only few periods could be digitally synthesized. We propose to use the focused ion beam (FIB) nanotechnology for waveguide OE fabrication [16]. The possibility of FIB obtaining periodic structures with high aspect ratio was demonstrated. We used the Strata[TM] FIB 201 machine with $Ga^+$-ion beam. The minimum diameter of the focused ion beam was 7 nm. We used two options. According to the first one, the periodic structure was directly fabricated on a polymer layer. For the second option, the periodic structure was firstly fabricated on a Si plate and then it was replicated on UV-cured polymer. The polymer samples were fabricated by V.I. Sokolov (Institute on Laser and Information Technologies of RAS, Troitsk, Russia). The periodic structures fabricated on polymer and silicon are shown on figures 4a and 4b respectively. The photos were produced at FIB by secondary electrons imaging. To see the grating profile, a special secant cavity normal to its grooves was produced. The photos were imaged at a tilt angle of 45°. The periodic structures with groove depth to width ratio about 20 were obtained. The groove width was about 200 nm near the top.

The achromatic waveguide OE can be used for operating with ultrashort pulses, which have wide spectrum. Also it can be used for the parallel processing of signals with different frequencies. These both advantages are useful for increasing the speed of information processing. Possible element functions include radiation focusing, input/output coupling, and connection between planar waveguides and optical fibers.

## ACHROMATIC OPTICAL AND OPTOACOUSTICAL ELEMENTS AND SYSTEMS

Similar to the 2D case, 3D OE can be obtained not only by direct holographic recording but also by using digital holography methods when the OE shape is firstly calculated and then OE is fabricated by some technology. In [17] Stetson discussed a possibility of achromatic wavefront reconstruction by a hypothetical lone isophase surface if it is made infinitely thin and reflective. This popular science paper was of great importance due to the consequent works. In [18] Sheridon cited Stetson and noticed that shapes of adjacent surfaces of interference fringe maxima are very close to each other. This fact was used in following experiment. Sheridon recorded a hologram in reflected geometry on a photoresist. After the photoresist treatment, its surface shape looked similar to the blazed grating. Each groove of this grating is a part of some interference fringe maximum surface. The adjacent groove relates to the adjacent surface of interference fringe maximum. A relief jump between adjacent grooves corresponds to $2\pi$ phase jump. One year later a computer generated OE called kinoform, which was similar to the Sheridon's blazed hologram was proposed [19]. Later, deep kinoforms (also called multiorder or harmonic diffractive lenses) with phase jump equal to $2\pi N$ (where $N$ is integer >1) were developed [20-24]. However, the blazed holograms and kinoforms are not achromatic. It is obvious that $2\pi$ or $2\pi N$ jumps of the surface relief correspond to the certain wavelength. To make OE achromatic we need to exclude the phase jumps. OE achromatically reconstructing some wavefront should be optically equivalent to one of interference fringe maximum surfaces when this wavefront interferes with some reference wave. The shape of this surface can be calculated or experimentally measured. The fabricated OE can be reflecting or transmitting. In the simplest case the surface of reflective OE coincides with the shape of interference fringe maximum surface. For transition from reflection OE to its transmitting equivalent its surface coordinates in direction of the optical axis should be multiplied by 2/(n-1) (where n is the material refractive index). The technology of OE fabrication depends on its shape and accuracy requirements [3]. However, if the amplitude of the OE relief depth variation is too big for technology of its fabrication some of following options can be used.

As the first option, a phase jump compensation is suggested [3, 25]. An optical system shown in figure 5a consists of a blazed element (BE) and a stepped element (SE) situated closely to each other. Similar to kinoforms the surface of the BE is divided into zones. The relief depth varies continuously within every zone and has jumps on the borders between neighbor zones. In contrary to kinoforms in BE the jump value is not connected to the wavelength. The working surface of each step of the SE represents a part of a plane normal to the optical axis. Each step of the SE has the corresponding zone in the BE. These step and zone are situated oppositely to each other and have the same shape in the top view. The differences of the optical path lengths between the neighboring steps of the SE have equal absolute value and the opposite sign with the phase jumps on the border between the corresponding zones of the BE. Let's consider how the optical system shown in figure 5a works. Let's neglect inclined rays (due to refraction) inside the optical system and also assume that for all zones $l_i << D_i^2/\lambda$, where $l_i$ is the average distance between i-th zone of the BE and the corresponding step of the SE and $D_i$ is their width. First, let's consider the option when the BE is transmissive and the SE is reflective. Under these conditions the optical system works in the following way. At first, radiation passes through the BE obtaining some phase modulation. On the border between zones 1 and 2 the optical path length discontinuously changes by the value equal to $a(n_1 - n_0)$, where $a$ is the value of the relief depth jump on the this border, $n_1$ is the refraction index of the BE material, and $n_0$ is the refraction index the medium between the BE and the SE. Then, radiation reflects from the surface of the SE. The optical path length difference between radiation reflected from steps 3 and 4 corresponding to zones 1 and 2, respectively, is equal to $2hn_0$, where $h$ is a height of step 3. Then, radiation passes through the BE for the second time.

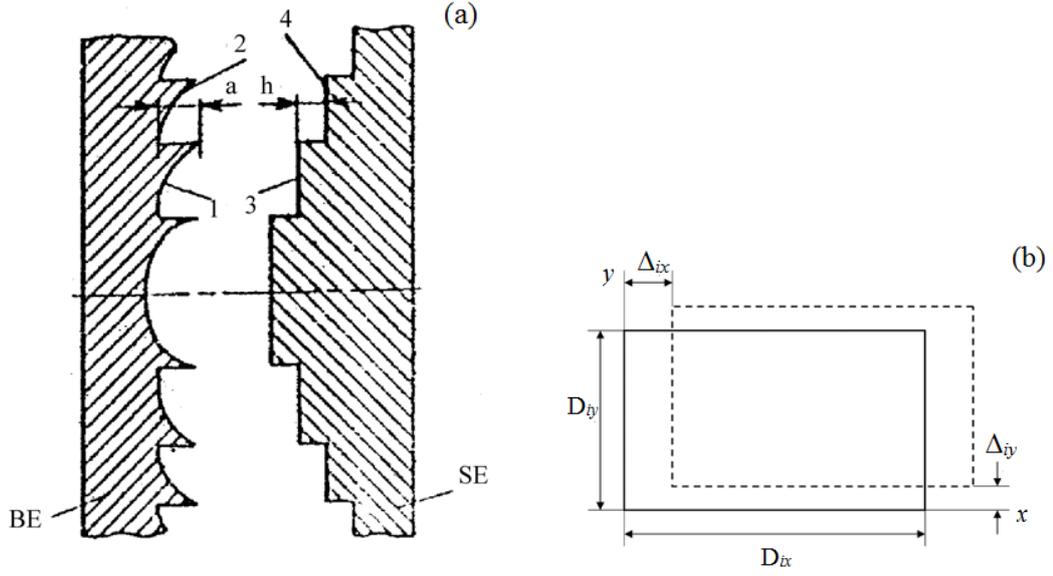

Figure 5. (a) Achromatic optiacal system containing BE and SE (1, 2 – zones of BE, 3, 4 – steps of SE) and (b) shift of the rays projections of i-th zone of BE relative to its corresponding step of SE.

We assume that the rays transmitted by each zone of the BE will be reflected only by the step of the SE corresponding to this zone and vice-versa. Thus the total optical paths difference between the rays leaving the optical system through zones 1 and 2 near the border between them is equal to $2a(n_1 - n_0) - 2hn_0$. In this case the condition of the phase jumps compensation mentioned above could be written as:

$$hn_0 = a(n_1 - n_0). \tag{8}$$

The optical system, for which (8) is satisfied, is equivalent to the OE with continuous variation of the relief depth. If material refractive index dispersion is neglected the optical system could be treated as achromatic. Note that refractive index dispersion is generally minor in comparison with dispersion due to diffraction by the periodic structure of a diffractive optical element [20].

For the other option, the BE of the optical system is reflective and the SE is transmitting. In this case the condition of the phase jumps compensation is the following:

$$h(n_2 - n_0) = an_0, \tag{9}$$

where $n_2$ is the refractive index of the SE material.

For the third option, both elements of the optical system are transmitting. In this case the condition of phase jumps compensation is the following:

$$h(n_2 - n_0) = a(n_1 - n_0), \tag{10}$$

If $n_1 = n_2$, $h = a$. The BE and SE could be produced on the opposite sides of the same substrate.

Earlier in (8-10) we have neglected inclined rays inside the optical system. Now, let's evaluate their impact. First, let's consider the case when both optical elements are transmitting. Due to the rays incline a part of radiation from the current zone of the BE does not hit its corresponding step of the SE. Let's treat this part of radiation as noise. Respectively, a part of radiation hitting the corresponding step is treated as a signal. Let's assume that the $z$ axis is directed along the optical axis and $\theta$ is an incline angle of rays with respect to $z$-axis. $\theta_x$ and $\theta_y$ are the projections of angle $\theta$ onto planes $xz$ and $yz$ respectively. For a quick evaluation of a signal to noise ratio (S/N) lower limit we will use a model in which all the rays coming from the current $i$-th zone of the BE have the same incline angle with the maximum values of projections $\theta_{ix}^{max}$ and $\theta_{iy}^{max}$. Let's assume that all zones have rectangular shape and $i$-th zone dimensions are equal to $D_{ix}$ and $D_{iy}$ in directions of related axes. Due to the rays incline the ray projection of $i$-th zone of the BE will be shifted in regards to its corresponding step of the SE by values $\Delta_{ix}$ and $\Delta_{iy}$ along related axes (figure 5b). The shifted ray projection of the current zone is shown in Figure 6 by a dotted line and its corresponding step is shown by a solid line. The shifts $\Delta_{ix}$ and $\Delta_{iy}$ are equal to:

$$\Delta_{ix} = l_i tan(q_{ix}^{max}) \sim l_i q_{ix}^{max},$$

$$\Delta_{iy} = l_i tan(q_{iy}^{max}) \sim l_i q_{iy}^{max}, \quad (11)$$

where $l_i$ is the average distance between the optical elements surfaces. We have substituted tangents for their arguments in (11). Let's treat the portion of the beam directed from the current zone into the corresponding step area as signal and the portion of the beam directed outside the one as noise. In this model the *S/N* lower limit is proportional to the ratio of the parts of the areas of the shifted zone projection, the first of which is inside and the other one is outside of the corresponding step (see figure 5b):

$$S/N > \sum_{i=1}^{N} \frac{(D_{ix}-\Delta_{ix})(D_{iy}-\Delta_{iy})}{D_{ix}D_{iy}-(D_{ix}-\Delta_{ix})(D_{iy}-\Delta_{iy})} = \sum_{i=1}^{N} \frac{(D_{ix}-l_i\theta_{ix}^{max})(D_{iy}-l_i\theta_{iy}^{max})}{D_{ix}D_{iy}-(D_{ix}-l_i\theta_{ix}^{max})(D_{iy}-l_i\theta_{iy}^{max})}$$

$$\approx \sum_{i=1}^{N} \left(\frac{D_{ix}D_{iy}}{l_i(D_{ix}\theta_{iy}^{max}+D_{iy}\theta_{ix}^{max})}-1\right) > \frac{1}{l\theta}\sum_{i=1}^{N}\frac{D_{ix}D_{iy}}{D_{ix}+D_{iy}}-N, \quad (12)$$

where N is a quantity of zones, $l$ and $\theta$ are the maximum values of $l_i$ and $\theta_{ix,y}^{max}$. We have neglected terms with the second power of $\theta_{ix,y}^{max}$ in (12) treating them as small parameters. For the cases when one of the optical elements is reflective $l$ should be doubled. It is proportional to the increase of the optical path length inside the system for this case. For some specific types of optical elements (for example, axially symmetric, etc.) a signal to noise ratio could be evaluated more exactly.

For the other option, the optical system has two OE not divided into zones [3]. The first OE has a strong optical power. We suppose that its surface can have deviations from the design shape due to the fabrication accuracy. These deviations are compensated by the second OE with a weak optical power. The week OE is fabricated using the results of additional measurements of the wavefront formed by the strong OE. Fabrication of the weak OE is significantly easier. The option when the strong OE is reflecting and the week OE is transmitting is here preferred because for this case the system chromatic aberrations due to the material dispersion of transmitting OE would be minimal.

Another option was proposed for the special type of OE called focusators [26]. According this option, various zones of OE focus parts of the image into separate areas of the focusing domain. So, here each zone of the focusator is operating as an independent GO OE.

The focusing optoacoustical element proposed in [27] is operating as follows. The laser pulse is directed onto the element surface. Acoustic wave is generated as a result of rapid light absorption near this surface. The patent [27] is based on the experimental results obtained by G.A. Askar'yan with collaborators [28, 29] in which a power ultrasonic pulse is generated at the spherical surface and focused into a focal point. Using power laser pulse or sequence of pulses and also metallized films bursting by electric discharge Askar'yan obtained the stress value about 1000 atm near focus. Instead of being spherical, the optoacoustical focusing element proposed in [27] has the complicated surface shape found by inverse solution for creating a local stress in the special shaped area inside the matter. Here the use of GO optical element is fundamentally because the acoustic pulse can have only one period [29]. The use of diffractive optical element for a single-cycle ultrasonic pulse is impossible. The accuracy requirement for the focusing optoacoustical element fabrication is significantly lower than for OE due to the difference between sound and light speeds. So, the focusing optoacoustical element can be fabricated by a numerically controlled machine. In [27] this technology was proposed for internal cracks inhibition in constructions by creating a plastic deformation area.

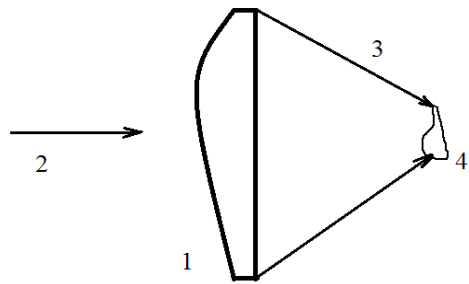

Figure 6. Optoacoustical focusing element: 1 - optoacoustical element, 2 – laser pulses, 3 – ultrasound pulses, 4 - stress area.

We suppose that it is possible to use this technique in other fields including surgical operation upon internals without cutting the upper body tissues. In principle, this method gives the possibility to reproduce the action of the scalpel or other surgical implements. Two options can be used for simulation the surgical implements movements inside the body. For the first one, the optoacoustical focusing element is moving along the patient's body. The element should be connected with the body through some immersion. For the second option, the optoacoustical focusing element is elongated along the way of the scalpel movement and the laser beam is moving along the element. Also, we suppose that this technique can be used for simulation of a bandage to stop internal bleeding.

## TEMPORAL RECONSTRUCTION

Additional effects arise when ultra-short pulses are used not only for hologram recording but also for reconstructing [30]. The reconstruction of the pulse temporal structure is among them. The following simple example shows how it works (figure 7). Suppose that the duration of the reference pulse is significantly less than the object pulse duration. So, we will consider the reference pulse as a $\delta$-pulse. Let the object beam consists of two $\delta$-pulses delayed with respect to each other by $\tau$. We suppose that the reflection geometry of recording is use. In that case the pulses time-space intersection consists of two parallel planes separated by a distance $\tau c/2$, where c is speed of light. Let the reconstruction pulse be identical to the reference one. It will be reflected sequentially from one plane and then from the other. The time delay between two reflected pulses will be equal to $\tau$. So, the time delay was reconstructed by simple GO reflection.

If the reconstructing wave is conjugated with the reference wave during recording the time-reversed pulse will be obtained. In our example shown on figure 7 this relates to the reconstructing $\delta$-pulse propagating towards the recording structure from the right side.

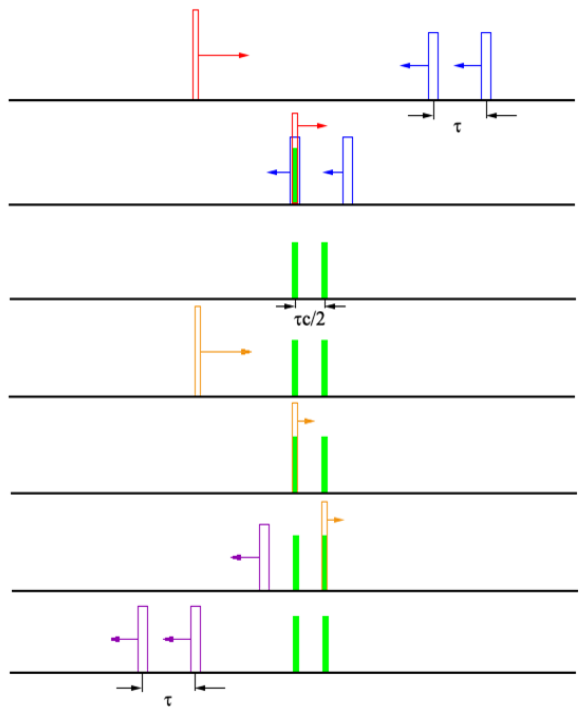

Figure 7. Temporal light pulse reconstruction

The several types of holographic temporal reconstruction are well known [31, 32]. We think that in our example we have another mechanism of reconstruction. Possibly, it is close to the mechanism of light pulse shaping by Bragg grating [33]. However, here we operate in time-domain instead of spectral-domain.

## CONCLUSION

The difference between the GO mechanism of wavefront reconstruction and the holographic mechanism was demonstrated. The GO wavefront reconstruction was obtained by using ultrashort laser pulses for recording in volume medium and in the planar optical waveguide. Several types of achromatic optical elements (OE) and systems were discussed. The optoacoustic focusing element, which can create a great local stress in a special shaped area inside the matter, was proposed. The possibility of the ultrashort pulse temporal reconstruction was demonstrated.

## ACKNOWLEDGMENTS

The work was carried out within the framework of the state task. I thank G. Chernov (Universidad de Sonora, Hermosillo, Sonora, México) for his help with language editing.